\documentclass[12pt]{article}
\usepackage{epsfig, amssymb}
\usepackage{graphicx,psfrag} 
\newcommand{\be}{\begin{equation}}
\newcommand{\ee}{\end{equation}}
\newcommand{\bea}{\begin{eqnarray}}
\newcommand{\eea}{\end{eqnarray}}
\newcommand{\bA}{\begin{array}}
\newcommand{\eA}{\end{array}}
\newcommand{\bc}{\begin{center}}
\newcommand{\ec}{\end{center}}
\newcommand{\al}{\alpha}

\newcommand{\ra}{\rightarrow}
\newcommand{\del}{\partial}

\newcommand{\ie}{{\it i.e.}}
\newcommand{\eg}{{\it e.g.}}

\newcommand{\Nf}{${\cal N}{=}4$}

\def\BC{{\mathbb C}}
\def\BP{{\mathbb P}}

\begin{document}

\begin{titlepage}
\vspace{30mm}

\bc

\hfill  {MIT-CTP/4151} \\
\hfill 
\\         [22mm]

{\huge {\bf Lifshitz spacetimes from \\ [2mm]
$AdS$ null and cosmological solutions}}
\vspace{16mm}

{\large Koushik Balasubramanian$^a$ and K.~Narayan$^b$} \\
\vspace{3mm}
{$^a$\small \it Center for Theoretical Physics, MIT, \\}
{\small \it Cambridge, Massachusetts 02139, USA.\\}
\vspace{3mm}
{$^b$\small \it Chennai Mathematical Institute, \\}
{\small \it SIPCOT IT Park, Padur PO, Siruseri 603103, India.\\}

\ec
\medskip
\vspace{30mm}

\begin{abstract}
We describe solutions of 10-dimensional supergravity comprising null 
deformations of $AdS_5\times S^5$ with a scalar field, which have 
$z=2$ Lifshitz symmetries. The bulk Lifshitz geometry in 
$3+1$-dimensions arises by dimensional reduction of these solutions. 
The dual field theory in this case is a deformation of the \Nf\ super 
Yang-Mills theory. We discuss the holographic 2-point function of 
operators dual to bulk scalars. We further describe time-dependent 
(cosmological) solutions which have anisotropic Lifshitz scaling 
symmetries. We also discuss deformations of $AdS\times X$ in 
11-dimensional supergravity, which are somewhat similar to the 
solutions above. In some cases here, we expect the field theory duals 
to be deformations of the Chern-Simons theories on M2-branes stacked 
at singularities.
\end{abstract}

\end{titlepage}

\newpage 
{\footnotesize
\begin{tableofcontents}
\end{tableofcontents}
}

\vspace{2mm}

\section{Introduction}

It is of interest to explore the space of physical systems that
constructions in string theory can (approximately) model, in particular 
containing some key qualitative features of the physical systems. 
In this light, the recent holographic discussions of nonrelativistic 
systems \eg\ \cite{Son:2008ye,Balasubramanian:2008dm,
Maldacena:2008wh,Herzog:2008wg,Adams:2008wt,Goldberger,Barbon,KLM,
marikataylor,hartnollSch,gauntlettSch,bobevkundupilch} with a view towards 
condensed matter physics are promising. 
In this paper, we discuss Lifshitz fixed points from a holographic 
point of view: see \eg\ \cite{horava1,horava2,danielsson,adamsmaloney,
takayanagiLifsh32,wadia,dasmurthy,TTnogo,mcgreevyBalasubLifBH,trivediLifBH,
HPST,danielsson2,faulkpolch,kachruLifQHE,danielsson3,koroteev}, for 
related work on systems with Lifshitz symmetries.

Various condensed matter systems admit descriptions in terms of Lifshitz 
fixed points, with dynamical exponent $z$ given by the anisotropic 
scaling\  $t\ra \lambda^z t,\ x^i \ra \lambda x^i$. A Landau-Ginzburg 
description for such theories with $z=2$ has the effective action\ 
$S = \int dt d^dx\ ((\del_t\varphi)^2 - \kappa (\nabla^2\varphi)^2)$. 
These theories, discussed early on in \cite{hornreich,grinstein}, arise 
in dimer models \eg\ \cite{ardfendfradkin}, representing universality 
classes of dimer statistical systems \cite{henley}, or as representing 
certain phases of systems with antiferromagnetic interactions as well 
as in models of liquid crystals \cite{chaikinluben}. It was argued in 
\cite{ardfendfradkin} that the equal time correlation functions of 
a (2+1)-dim Lifshitz theory are identical to the correlators of an 
appropriate Euclidean 2-dim conformal field theory. Further it was 
discussed in \cite{ashvinSenthilLifsh} in the context of quantum 
critical points that finite temperature equal-time correlators of 
these theories exhibit ultra-locality in space.

Holographic duals of Lifshitz-like theories were studied in \cite{KLM}. 
They found that the following metric provides a geometric realization 
of the symmetries of Lifshitz-like theories (with $z$ as the dynamical 
exponent):
\be
ds^2 = -{dt^2\over r^{2z}} + {dx_i^2 + dr^2 \over r^2}\ ,
\ee
where ${\vec x}\equiv x_i$ denotes a $d-$dimensional spatial vector. 
In the case $d=2$, this metric is a classical solution of the 
following action:
\be\label{KLMaction}
S = {1\over 2} \int d^4 x \left( R - 2 \Lambda\right) - {1\over 2} \int 
\left( F_{(2)} \wedge \star F_{(2)} + F_{(3)} \wedge \star F_{(3)}\right) 
- c \int B_{(2)}\wedge F_{(2)}\ ,
\ee
where, $ F_{(2)}= dA_{(1)}$,  $ F_{(3)}= dB_{(2)}$ and $\Lambda$ is the 
4-dimensional cosmological constant. By dualizing the $B_{(2)}$-field, 
one obtains a scalar $\varphi$, with the $B_2\wedge F_2$ term recast 
as\ $-\sqrt{-g} \del^\mu\varphi A_\mu$: integrating out the scalar then 
gives a term $A^2$, \ie\ a mass term for the gauge field. In other 
words, this system of fluxes can be rewritten as a massive 4-dim 
gauge field \cite{marikataylor,mcgreevyBalasubLifBH}, with profile\ 
$A\sim {dt\over r^z}$\ in the Lifshitz background.

Recently, some obstacles in finding a string construction of such
theories were pointed out in \cite{TTnogo}. They showed, with
reasonable ansatze for the fluxes, that it is not possible to have a
classical solution of massive type IIA supergravity/M-theory of the
form $Li_{4} \times M_{6}~(\mbox{or }M_{7})$\footnote{Note however 
\cite{takayanagiLifsh32}, which uses intersecting D3-D7 branes to
construct $z={3\over 2}$ Lifshitz spacetimes that are anistropic and
in addition have a nontrivial dilaton that breaks this symmetry. Note 
also \cite{marikataylor,danielsson2}, which construct Lifshitz-like 
solutions with a scalar having a radial profile. See also 
\cite{koroteev} which describes anisotropic Lifshitz-like solutions 
with anisotropic matter.}. This was shown to be true even when the 
product contains warp factors. To the best of our knowledge, solutions 
of 10- or 11-dimensional supergravity with Lifshitz symmetries have not 
yet been constructed. However, some ways of overcoming these obstacles
were outlined in \cite{HPST}.

In this note we suggest alternative constructions, with explicit
solutions of supergravities which have $z=2$ Lifshitz symmetries. 
Lifshitz theories with dynamical exponent $z=2$ are closely related to
Galilean invariant CFTs (Schr\"odinger invariant theories). Note that
Lifshitz theories have only non-relativistic scale invariance: these
theories are not Galilean invariant. These theories do not have a
conserved particle number unlike Galilean invariant theories. This
suggests that Lifshitz invariant theories can be constructed by
explicitly breaking Galilean invariance in Schr\"odinger invariant
theories.
We recall that holographic descriptions of Galilean invariant CFTs 
(with Schr\"odinger symmetry) were proposed in \cite{Son:2008ye,
Balasubramanian:2008dm}: they can be embedded in string theory 
\cite{Maldacena:2008wh,Herzog:2008wg,Adams:2008wt,hartnollSch,
gauntlettSch,bobevkundupilch}. In this 
description, the particle number symmetry is geometrically realized as 
an isometry of a circle (denoted by $x^+$). The geometry in this 
description has some resemblance to $AdS$ in lightcone coordinates, 
with one of the lightcone directions compactified. In fact, $AdS$ in 
light cone coordinates (with a compact lightcone direction) has the 
symmetries of the Schr\"odinger group \cite{Goldberger, Barbon}. 
With a view to breaking the Schrodinger symmetry to a Lifshitz one, 
the shift symmetry along $x^{+}$ direction can be broken in many ways. 
For instance, adding backreacting branes (anti-branes) that are
localized along this compact direction breaks this shift symmetry
explicitly resulting in a geometry with Lifshitz symmetries. 

Our construction in this paper describes a possibly simpler way of 
breaking this shift symmetry by turning on a scalar field periodic in 
$x^{+}$ (with period determined by the radius of the $x^{+}$ direction). 
A scalar field with profile $\Phi(x^+)$ breaks the shift symmetry 
(asymptotic) along $x^+$ direction. Such solutions of supergravity have 
already been studied in the literature with a view to understanding 
cosmological singularities in AdS/CFT \cite{dmnt1, dmnt2,adnt,adnnt}. 
We will review relevant aspects of these in the next section (sec. 2), 
but for now we describe some essential features of our proposed 
holographic system exhibiting $z=2$ Lifshitz symmetry. The spacetimes 
we deal with are solutions of 10- or 11-dim supergravity comprising 
deformations of $AdS\times X$, alongwith a scalar $\Phi(x^+)$, the 
AdS-deformed metric being
\be\label{metPBH}
ds^2 = {1\over w^2} [-2dx^+dx^- + dx_i^2 + \gamma (\Phi')^2 w^2 (dx^+)^2]
+ {dw^2\over w^2} + d\Omega_S^2\ ,
\ee
with\ $\Phi'\equiv {d\Phi\over dx^+}$\ . 
The constant $\gamma$ is\ $\gamma={1\over 4}$ for $AdS_5$ and 
$\gamma={1\over 2}$ for $AdS_4$, with the $x_i$ ranging over $1,2$ and 
$1$ for $AdS_5$ and $AdS_4$ respectively (the $d\Omega_S^2$ is the 
metric for $S^5$ or $X^7$ respectively, with $X^7$ being some 
Sasaki-Einstein 7-manifold). We regard $x^-$ as the time direction here, 
$x^+$ being a compact direction. We will discuss this metric in greater 
detail in the next section: there we will also describe a more general 
context that these solutions (and others discussed later) will naturally 
arise from.

It can be checked that these spacetimes (\ref{metPBH}) along with 
the scalar $\Phi$ and appropriate 5-form (or 4-form) field strength 
are solutions to the 10-dim (or 11-dim) supergravity equations. For 
instance, there is no $S^5$ or $X^7$ dependence and the resulting 5- 
or 4-dim system, with an effective cosmological constant from the flux, 
solves the equations\ $R_{MN}=-d g_{MN} + {1\over 2} \del_M\Phi\del_N\Phi$, 
with $d=4,3,$ for $AdS_{d+1}$, being the 5- or 4-dim effective 
cosmological constant.\\
The spacetime (\ref{metPBH}) exhibits the following symmetries: 
translations in $x_i,x^-\equiv t$ (time), rotations in $x_i$ and a 
$z=2$ scaling\ 
$x^-\ra \lambda^2 x^- , x_i\ra \lambda x_i , w\ra \lambda w$\ ($x^+$ 
being compact does not scale). Possible Galilean boosts\ 
$x_i\ra x_i-v_ix^- ,\ x^+\ra x^+-{1\over 2} (2v_ix_i-v_i^2x^-)$,\ are 
broken by the $g_{++}\sim (\Phi')^2$ term. If $g_{++}=0$, this is 
essentially $AdS$ in lightcone coordinates and the system has a 
Schrodinger symmetry (as discussed in \eg\ \cite{Goldberger,Barbon,
Maldacena:2008wh}): note however that these are not Schrodinger 
spacetimes of the sort discussed in \cite{Son:2008ye,
Balasubramanian:2008dm,Maldacena:2008wh,Herzog:2008wg,Adams:2008wt,
hartnollSch,gauntlettSch,bobevkundupilch}. Similarly, in the present 
case with $g_{++}\neq 0$, there is no special conformal symmetry either. 
We discuss various aspects of this system in sec. 3 and sec. 4: this
includes a discussion of the dimensional reduction of these systems
and some aspects of the dual field theory (in part borrowing from
\cite{dmnt2}), which is the lightlike dimensional reduction, or DLCQ, 
along the $x^+$-direction of \Nf\ super Yang-Mills theory with a 
nontrivial gauge coupling\ $g_{YM}^2=e^{\Phi(x^+)}$. In particular we 
also discuss the holographic 2-point function of operators dual to bulk
scalars. Our equal-time holographic 2-point function in particular
recovers the spatial power-law dependence obtained in \cite{KLM}. It
is perhaps worth mentioning that the Lifshitz field theory here is an
interacting strongly coupled dimensionally reduced limit of the \Nf\
SYM theory, rather than a free Lifshitz theory.

Similarly we expect that the $AdS_4$-deformed solutions are dual to 
appropriate lightlike deformations of Chern-Simons theories arising 
on M2-branes stacked at appropriate supersymmetric singularities 
\cite{abjm,klebanov,terashima,jafferis,imamura,uedayama,lee,ms1,ms2,kleb2,
hanany}, dimensionally reduced along a compact direction.

In sec. 5, we discuss time dependent deformations of $AdS_5$ and 
$AdS_4$: in particular the asymmetric Kasner-like solutions also 
exhibit interesting (anisotropic) Lifshitz scaling symmetries, as we 
describe there. These solutions are qualitatively different from 
the null ones above (\ref{metPBH}), as we discuss. We also describe 
some aspects of the dual field theories.

In sec. 6, we describe a solution of 5-dimensional gravity with
negative cosmological constant and a massless complex scalar, that 
are similar to the null solutions (\ref{metPBH}) above: these upon 
dimensional reduction give rise to $2+1$-dim Lifshitz spacetimes.
This 5-dim solution can be uplifted to 11-dimensional supergravity.

Sec. 7 closes with a Discussion, while Appendix A provides some 
technical details for completeness.

\section{$AdS$ null and cosmological solutions}

The following solutions are discussed in \cite{dmnt1,dmnt2,adnt,adnnt} 
as cosmological generalizations of $AdS_5/CFT_4$.
The ten-dimensional Einstein frame metric, the scalar $\Phi$, and 
5-form flux are
\bea \label{geomscal}
&& ds^2=\frac{R^2}{r^2} ({\tilde g}_{\mu\nu} dx^\mu dx^\nu + dr^2) 
+ R^2 d\Omega_5^2 , \qquad  \Phi=\Phi(x^\mu)\ , \nonumber\\
&& \qquad\qquad\qquad\qquad F_{(5)}=R^4(\omega_5 + *_{10} \omega_5)\ ,
\eea
with $d\Omega_5^2$ being the volume element and $\omega_5$ being the 
volume form of the unit five sphere $S^5$. 
This is a solution of the ten dimensional Type IIB supergravity 
equations of motion as long as the four-dimensional metric, 
${\tilde g}_{\mu\nu}$, and the scalar $\Phi$, are only dependent on the 
four coordinates, $x^\mu, \mu =0,1,2,3$, and satisfy the conditions, 
\be\label{condns}
{\tilde R}_{\mu\nu} = \frac{1}{2} \del_\mu\Phi \del_\nu\Phi, \qquad
\del_\mu (\sqrt{-{\tilde g}}\ {\tilde g}^{\mu \nu} \del_\nu\Phi) = 0\ ,
\ee
where ${\tilde R}_{\mu\nu}$ is the Ricci curvature of the metric 
${\tilde g}_{\mu\nu}$: these are equations governing 4-dim Einstein 
dilaton gravity.

The scalar $\Phi$ can be taken to be the dilaton with $e^\Phi$ then
being the string coupling. As described in Appendix A, more general 
solutions exist where the $S^5$ is replaced by the base of any 
Ricci-flat 6-dim space: in these cases, $\Phi$ can be taken to be some 
other scalar, \eg\ arising from the compactification.

Some details on these solutions that might be of relevance to the 
present context are reviewed in Appendix A.

We now specialise to null solutions where ${\tilde g}_{\mu\nu}$ and 
$\Phi$ are functions of only a lightlike variable $x^+$: if we further 
assume that ${\tilde g}_{\mu\nu}$ is conformally flat\ 
${\tilde g}_{\mu\nu}=e^{f(x^+)} \eta_{\mu\nu}$, the metric and dilaton 
become (setting the AdS radius $R=1$) for the $AdS_5$ case
\be\label{metconf5}
ds^2={1\over r^2} [e^{f(x^+)} (-2dx^+dy^- + dx_i^2) + dr^2] + d\Omega_5^2, 
\qquad \Phi=\Phi(x^+) 
\ee
(see also \cite{Chu:2006pa,linwen,adgot} for related work).
We use the variable $y^-$ for convenience, reserving $x^-$ for 
(\ref{metPBH}).
We will refer to the coordinate system in (\ref{metconf5}) as 
conformal coordinates in what follows.
The equations of motion in this case simplify drastically due to the 
lightlike nature of the solutions. The scalar equation of motion is 
automatically satisfied and the only nonzero Ricci component is 
$R_{++}$, giving\ $R_{++} = {1\over 2} (\del_+\Phi)^2$, \ie\ 
\be\label{eom5D}
R_{++} = {1\over 2} (f')^2 - f'' = {1\over 2} (\Phi')^2\ ,
\ee
with $\Phi'\equiv {d\Phi\over dx^+}\ ,\ f'={df\over dx^+}$\ .\ 
This is a single equation for two functions $f, \Phi$, so that this 
is a fairly general class of solutions with a function-worth of 
parameters: choosing a generic $\Phi$ gives an $e^f$. 
One has to be careful though, since an arbitrary $\Phi$ does not 
necessarily give an $e^f$ such that the pair is a sensible 
solution\footnote{For instance, in some related cosmological solutions 
and discussion in \cite{knNullws}, certain regulated versions of 
singular solutions do not necessarily obey $R_{++}>0$, which is 
essentially positivity of the energy density along null geodesics.}.
These solutions preserve half (lightcone) supersymmetry \cite{dmnt1}.

$AdS_4$ similarly admits generalizations of the solutions described 
above with the 11-dim metric and a scalar of the form\ 
$ds^2 = {R^2\over r^2} ({\tilde g}_{\mu\nu} dx^\mu dx^\nu + dr^2) 
+ R^2 d\Omega_{X^7}^2 , \ \Phi=\Phi(x^\mu)$. In this case, the scalar 
does not have any natural interpretation in the 11-dim theory directly: 
it arises instead from the 4-form flux after compactification on a 
7-manifold $X^7$ as we discuss in Appendix A.1. The 11-dim supergravity 
equations are satisfied if the conditions in (\ref{condns}) hold, the 
${\tilde R}_{\mu\nu}$ now being the Ricci tensor for the 3-dim metric 
${\tilde g}_{\mu\nu}$. Pure 3-dim gravity has no dynamics but the scalar 
drives the system giving rise to nontrivial dynamics. Consider now a 
3-dim metric conformal to flat 3-dim spacetime: the 11-dim metric then 
in conformal coordinates is
\be\label{metconf4}
ds^2 = {1\over r^2} [e^{f(x^+)}(-2dx^+dy^- + dx_i^2)+dr^2] + d\Omega_{X^7}^2\ ,
\qquad \Phi=\Phi(x^+)\ .
\ee
The Einstein equation becomes
\be\label{eom4D}
R_{++}={1\over 4} (f')^2-{1\over 2} f''={1\over 2} (\Phi')^2\ ,
\ee
of a form similar to the $AdS_5$ case.

Such a deformation, via a ${\tilde g}_{\mu\nu}$, could potentially lead 
to singularities on the Poincare horizon $r=0$. For instance in the 
$AdS_5$ case, we have\ $R_{\mu\nu\al\beta}={1\over r^2} {\tilde 
R}_{\mu\nu\al\beta} - 2 (g_{\mu\al}g_{\nu\beta}-g_{\mu\beta}g_{\nu\al}) ,\ 
R_{z\mu z\mu}=-{2\over r^2} g_{\mu\nu}$, using \eg\ \cite{Wald}, giving 
the curvature invariant\
$R_{ABCD} R^{ABCD}= r^4 {\tilde R}_{\mu\nu\al\beta} {\tilde R}^{\mu\nu\al\beta} 
+ O(r^0)$. Now for the null metrics in question here, 
${\tilde R}_{\mu\nu\al\beta} {\tilde R}^{\mu\nu\al\beta}$ vanishes, since 
the lightlike solutions admit no nonzero contraction. Thus the possible 
divergent $r^4$ term at the Poincare horizon $r\ra\infty$ is in fact 
absent. These null solutions are thus regular, except for possible 
singularities arising when $e^f$ vanishes, as in the context of 
cosmological singularities \cite{dmnt1,dmnt2,adnt,adnnt}: in that case, 
there were diverging tidal forces along null geodesics arising because 
the spacetime was essentially undergoing a crunch with $e^f$ vanishing. 
For our purposes here, $e^f$ and $\Phi$ will be regular functions of 
$x^+$, in which case we expect that these spacetimes are regular. It 
is however known that Lifshitz spacetimes have diverging tidal forces 
\cite{KLM}\ (see also \cite{Blau} which describes various geometric 
properties of the Schrodinger and Lifshitz spacetimes). It would seem 
that the singularities of the Lifshitz geometry then arise from the 
process of dimensional reduction of the above spacetimes (discussed 
in the next section).

In many cases, it is possible to find new coordinates such that boundary 
metric\ $ds_4^2=\lim_{r\ra 0} r^2 ds_5^2$\ ($AdS_5$)\ or\ 
$ds_3^2=\lim_{r\ra 0} r^2 ds_4^2$\ ($AdS_4$) is flat, at least as 
an expansion about the $r=0$ boundary, if not exactly: this was studied 
for $AdS_5$ null cosmologies in \cite{adnt}). These are {\it 
Penrose-Brown-Henneaux (PBH)} transformations, a subset of bulk 
diffeomorphisms leaving the metric invariant (in Fefferman-Graham form), 
and acting as a Weyl transformation on the boundary.

The coordinate transformation\  $w=r e^{-f/2} , \ x^-=y^--{w^2 f'\over 4} $,
\ recasts these spacetimes (\ref{metconf5}), (\ref{metconf4}), in the 
form (\ref{metPBH}), reproduced here,
\be\label{metPBH2}
ds^2 = {1\over w^2} [-2dx^+dx^- + dx_i^2 + \gamma (\Phi')^2 w^2 (dx^+)^2]
+ {dw^2\over w^2} + d\Omega_S^2\ , 
\ee
using the equation of motion (\ref{eom5D}) or (\ref{eom4D}) for 
these solutions, with\ $\gamma={1\over 4}$ for $AdS_5$ and 
$\gamma={1\over 2}$ for $AdS_4$. Likewise, the $x_i$ range over $1,2$ 
and $1$ for $AdS_5$ and $AdS_4$ respectively. We refer to this metric 
as written in PBH coordinates.
In this lightlike case, this is an {\it exact} PBH 
transformation\footnote{For $t$-dependent solutions, an exact PBH 
transformation is difficult to find in general, and one instead takes 
recourse to an expansion about the boundary \cite{adnt}.}.
Now the boundary at $w=0$ is manifestly flat 4D or 3D Minkowski 
spacetime, for the $AdS_5$ or $AdS_4$ cases respectively.
With any infinitesimal regulator however, the regulated boundary 
$r=\epsilon$ is distinct from $w=\epsilon$, \ie\ the holographic screens 
are distinct, although in the same conformal class.\\
Note that these are not normalizable deformations: \eg\ in the 
$AdS_5$-deformed case, those would correspond to deformations where 
$w^2 g_{++}\sim w^4$.

In the next section, we study the dimensional reduction of these systems 
(\ref{metPBH}) (\ref{metPBH2}) with a view to realizing spacetimes with 
Lifshitz symmetries \cite{KLM} as a Kaluza-Klein reduction in one lower 
dimension.

\section{Dimensional reduction to Lifshitz spacetimes}

In the cosmological singularities context \cite{dmnt1,dmnt2,adnt,adnnt},
$x^+$ was regarded as a lightcone time coordinate, working in the
conformal coordinate system (\ref{metconf5}): this introduces
nontrivial lightcone time dependence into the system. From the dual
gauge theory point of view, this makes the gauge coupling
$g_{YM}^2=e^{\Phi(x^+)}$ time dependent. Note that the boundary metric
is either flat (in the PBH coordinates) or conformally flat (in the
conformal coordinates): thus $x^+$ can be regarded equally well as a
lightcone time or space variable in the boundary theory.
In the bulk, although the worldsheet string is difficult to understand
technically, it is natural to study string propagation on such
spacetimes by fixing lightcone gauge as\ $\tau=x^+$, where $\tau$ is
worldsheet time. In a sense, this has some parallels (and also some
key differences) with the investigations of strings in plane wave
spacetimes (see \eg\ \cite{mukundPP} for discussions of global
properties and time-functions in plane wave spacetimes).\\
However, regarding $x^+$ as a time coordinate might appear problematic
in the PBH coordinate system (\ref{metPBH}), (\ref{metPBH2}), since 
$g_{++}=\gamma (\Phi')^2>0$, implying $\del_+$ is a spacelike vector. 
Strictly speaking, the $x^+=const$ surfaces are null surfaces since
their normal $dx^+$ is null, noting that $g^{++}=0$, while $x^-=const$
surfaces are spacelike, given that $g^{--}<0$, suggesting again that
$x^-$ behaves like a time coordinate.

Now if $x^+$ represents a compact dimension, the discussion above 
needs to be qualified. Specifically the case $g_{++}<0$ with $x^+$ 
treated as the time coordinate signals the presence of a closed timelike 
curve if $x^+$ is a compact dimension. In the present context, we have 
$g_{++}\sim \gamma(\Phi')^2>0$, and it is sensible to compactify $x^+$ on 
a spacelike circle. That is, we consider $x^-$ to be the time coordinate. 
In this case, these are spacetimes with no $x^-$ dependence, \ie\ with 
time translation invariance. The scalar field must be a periodic 
nonsingular function $\Phi(x^+)$. A periodic $\Phi$ varying nontrivially 
over the compact $x^+$-direction has $\Phi'=0$ at isolated $x^+$-values: 
this however is not problematic, since it will turn out that $\Phi'$
essentially disappears.  At long wavelengths compared with the size of
the $x^+$-circle, this gives an effective bulk 4-dim or 3-dim
spacetime. The Kaluza-Klein compactification is natural and manifest
in the PBH coordinates (\ref{metPBH}) (\ref{metPBH2}): it can be 
performed in the standard way by writing\  
$ds^2 = g_{mn} dx^m dx^n 
= G_{\mu\nu} dx^\mu dx^\nu + G_{dd} (x^d + A_\mu dx^\mu)^2 $.\  
Then the $\{g_{++},g_{+-}\}$-terms can be rewritten as
\bea
{1\over w^2} \left( \gamma w^2 (\Phi')^2 (dx^+)^2 - 2dx^+dx^- \right)
= \gamma (\Phi')^2 \left(dx^+ - {dx^-\over \gamma w^2 (\Phi')^2} \right)^2 
- {(dx^-)^2\over \gamma w^4 (\Phi')^2} \ . \nonumber
\eea
Thus the effective 4D or 3D metric, for $AdS_5$ or $AdS_4$ respectively, 
after compactifying on $x^+$ naively becomes
\be\label{metlifsh}
ds^2 = - {(dx^-)^2\over \gamma w^4 (\Phi')^2} + {dx_i^2\over w^2} + 
{dw^2\over w^2}\ ,
\ee
where $\gamma$ and the range of $x_i$ have been defined after 
(\ref{metPBH}). Apart from the annoying factor of $(\Phi')^2$ which 
disappears as we will see below, these are thus spacetimes which 
exhibit a Lifshitz-like scaling with exponent $z=2$, \ie\ 
\be
x^-\equiv t \ra \lambda^2 t ,\ \quad\  x_i\ra \lambda x_i ,\ \quad\
w\ra \lambda w\ .
\ee
The $z=2$ Lifshitz scaling can also in fact be seen in the metric written 
in conformal coordinates (\ref{metconf5}): taking the compact coordinate 
$x^+$ to not scale, we see the scaling\ $y^-\sim w^2$. This is also 
consistent with the conformal-PBH coordinate transformation relation\ 
$y^-=x^-+{w^2 f'\over 4} \sim \lambda^2 y^-$.\ Likewise, the presence of 
the conformal factor $e^{f(x^+)}$ breaks boost invariance.

These Lifshitz spacetimes are likely to not have any supersymmetry.
However the null solutions described previously in fact do preserve 
some fraction of lightcone supersymmetry. Our belief is that the 
dimensional reduction along the $x^+$-direction breaks the 
lightcone supersymmetry completely.

Note that the nontrivial dependence on the $x^+$-direction through the 
$g_{++}\sim (\Phi')^2$ term breaks the Galilean boost invariance,\ 
$x_i\ra x_i-v_ix^- ,\ x^+\ra x^+-{1\over 2} (2v_ix_i-v_i^2x^-)$. If 
$\Phi'=0$, then $g_{++}=0$, boost invariance reappears, and the system 
has a larger Schrodinger symmetry.\\
If $x^+$ is noncompact, these systems admit a lightlike scaling symmetry\ 
$x^+\ra \lambda x^+ ,\ x^-\ra {1\over\lambda} x^-,\ Q\ra {Q\over\lambda}$ ,\ 
where the parameter $Q$ appears in the combination $Qx^+$ in any 
function of $x^+$, \eg\ $e^{f(x^+)}=e^{f(Qx^+)}$. This can be used to 
fix the parameter, say as $Q=1$. The compactification of the 
$x^+$-dimension makes the system nonrelativistic, the compactification 
size becoming a physical (inverse) mass parameter. This lightlike 
scaling then is not a physical symmetry anymore, since it changes the 
physical parameters of the compactified nonrelativistic theory.
The PBH coordinate system allows a natural interpretation to the 
compactification process: technically, this admits a natural 
Kaluza-Klein reduction by compactification on $x^+$.

\subsection{Dimensional reduction, more rigorously}

Consider a 5-dim metric of the form
\be\label{metDimRed}
ds^2 = -N^2(x^+) K^2(s^i) dt^2 + {1\over N^2(x^+)} (dx^+ + N^2(x^+) A)^2 
+ {1\over w^2} (ds^i)^2\ ,
\ee
where $N(x^+)$ governs the metric component $g_{++}$, with $A$ being the 
Kaluza-Klein gauge field, and $s^i=x^i,w$ ($x^i\equiv x^1,x^2$). 
We have identified $t$ as $x^-$ earlier: the metric has no $t$-dependence.
Define vielbeins\footnote{The metric in component form is
\bea
ds^2 = -N^2(x^+) K^2(s^i) (1-A_0^2(s^i)) dt^2 + {(dx^+)^2\over N^2(x^+)} 
+ 2A_0(s^i) K(s^i) dx^+ dt + g_{ij} ds^ids^j\ , \nonumber
\eea }
\be
{\bar e}^0 = N e^0 = N K dt\ , \quad 
{\bar e}^+ = {1\over N} (dx^+ + N^2 A_0 K dt)\ ,\quad 
{\bar e}^i = e^i = {1\over w} ds^i\ ,
\ee
where ${\bar e}^\mu$ are vielbeins in the 5-dim metric, while $e^\mu$ 
are those of the lower dimensional metric: these satisfy\ 
$ds^2 = \eta_{MN} {\bar e}^M {\bar e}^N$.\  
We take the Kaluza-Klein gauge field defined by the 1-form\ 
$A=A_0 e^0 = {A_0\over N} {\bar e}^0$ to comprise purely a scalar 
potential with solely electric field strength, defined as\ 
$dA = {1\over 2} F_{0i} e^0\wedge e^i$, in terms of the vielbeins $e^\mu$ 
of the lower dimensional spacetime. The field strength is related to 
the gauge field as\ $F_{0i} = -2w (\del_iA_0 + A_0 {\del_i K\over K} )$.
This is thus a ``minimal'' metric family that contains the $AdS_5$ null 
solution we have been discussing above.

Furthermore, we obtain a null-type metric of the form we have discussed 
earlier if we set\ $g_{tt}=-N^2 K^2 (1 - A_0^2)=0$, \ie\ $A_0^2=1$:\ 
comparing with the earlier metric (\ref{metPBH}) (\ref{metPBH2}), we see 
that\ $N={1\over\sqrt{\gamma} \Phi'}\ ,\ K={1\over w^2}$ . Dimensionally, 
we have\ $[N]=L, [K]=M^2, [A_0]=0$, and $[e^A]=0$, \ie\ all vielbeins are 
dimensionless, consistent with the fact that the metric is dimensionless 
in units where $R_{AdS}=1$ (the lhs is actually ${ds^2\over R_{AdS}^2}$). 
With this simplified ansatz however, it is difficult to separate the 
gauge field parts of the system from the lower dimensional metric per se: 
in other words, it is desirable to retain $K(s^i)$ and $A_0(s^i)$ 
separately towards understanding the lower dimensional effective action 
better. 

We define the spin connection $\omega^a{_b}$ via the relations\ 
$d{\bar e}^a = - \omega^a{_b}\wedge {\bar e}^b$.\  We have\
(note \eg\ $\omega^0{_+}=-\omega_{0+}=\omega_{+0}=\omega^+{_0}$)
\bea
d{\bar e}^0 &=& - \omega^0{_+}\wedge {\bar e}^+ - 
\omega^0{_i}\wedge {\bar e}^i = 
{w \del_i K\over K} {\bar e}^i\wedge {\bar e}^0 + 
N' {\bar e}^+\wedge {\bar e}^0\ , \nonumber\\
d{\bar e}^+ &=& - \omega^+{_0}\wedge {\bar e}^0 - 
\omega^+{_i}\wedge {\bar e}^i = {1\over 2} F_{0i} {\bar e}^0\wedge {\bar e}^i 
+ N' A_0 {\bar e}^+\wedge {\bar e}^0\ ,  \\
d{\bar e}^i &=& -\omega^i{_0}\wedge {\bar e}^0 
- \omega^i{_+}\wedge {\bar e}^+ - \omega^i{_j}\wedge {\bar e}^j =
- {\bar e}^w\wedge {\bar e}^i\ ,\nonumber
\eea
and the spin connection becomes
\bea
\omega^0{_+} = \omega^+{_0} = N' {\bar e}^0 - N' A_0 {\bar e}^+ + 
{1\over 4} F_{0i} {\bar e}^i\ , \quad &&
\omega^i{_+} = -\omega^+{_i} = {1\over 4} F_{0i} {\bar e}^0\ ,\nonumber\\
\omega^0{_i} = \omega^i{_0} = {w \del_i K\over K} {\bar e}^0 
+ {1\over 4} F_{0i} {\bar e}^+\ , \quad &&
\omega^i{_w} = -\omega^w{_i} = - {\bar e}^i \ .
\eea
The curvature 2-forms are calculated using\ 
$R^a{_b} = d\omega^a{_b} + \omega^a{_c}\wedge \omega^c{_b} 
= R^a{_{bcd}} {\bar e}^c\wedge {\bar e}^d$.
The relevant Riemann tensor components are
\bea
R^0{_{+0+}} = -(NN''+{N'}^2) (1-A_0^2) - {1\over 16} F_{0i}^2\ , \quad
R^i{_{+i+}} = - {1\over 16} F_{0i}^2\ , \quad 
R^i{_{jij}} = -1 = R^i{_{wiw}} \ \ [i,j\neq w] , \nonumber\\
R^0{_{i0i}} = {w\del_wK\over K} - {w^2\del_i^2K\over K} 
+ {1\over 8} F_{0i}^2\ \ [i\neq w]\ , \qquad 
R^0{_{w0w}} = -{w\del_wK\over K} - {w^2\del_w^2K\over K} 
+ {1\over 8} F_{0w}^2\ .\ \qquad\ \
\eea
The metric determinant is $-g=-{K^2\over w^6}$, and the Ricci scalar for 
this metric is 
\bea\label{Ricci5dim}
R^{(5)} &=& {1\over 2K^2}  \Biggl[ -4 (NN'' + (N')^2) K^2 - 12 K^2 
- 4 w^2 K \del_i^2 K + 4 w K\del_w K \nonumber\\
&& \ + \left[ 4 (N N'' + (N')^2) K^2 + w^2 (\del_i K)^2 \right] A_0^2 
+ 2 w^2 K A_0 \del_i A_0 \del_i K + w^2 K^2 (\del_i A_0)^2 \Biggr] , \nonumber\\
&=&  -2(NN'' + (N')^2) - \left[{2\over K} (w^2 \del_i^2 K - w \del_w K + 3K)
\right] + {1\over 8} F_{0i}^2 + 2 (NN'' + (N')^2) A_0^2\ .\qquad
\eea
(Numerical output corroborates this.)
This higher dimensional Ricci scalar expanded in terms of the lower
dimensional modes essentially gives the lower dimensional effective
action on wavelengths long compared with the size of the compact
dimension. Note that if there was no nontrivial $x^+$-dependence in this
system, this would be the conventional Kaluza-Klein reduction with the
lower dimensional fields (metric, massless gauge field and scalar) being
independent of the compact dimension. The scalar $g_{++}={1\over N^2(x^+)}$ 
in this case is of a restrictive form, which therefore reflects in its 
lower dimensional kinetic term being a total derivative\ $\del_+(NN')$.

The form of $R^{(4)}$ appearing here suggests that the lower dimensional 
spacetime is in fact of the form
\be\label{met4Lif}
ds^2 = -K^2(s^i) dt^2 + {1\over w^2} {ds^i}^2 \quad\ \Rightarrow\ \quad\
R^{(4)} = -{2\over K} \Big( w^2 \del_i^2 K - w \del_w K + 3K \Big)\ .
\ee
Note that the $N(x^+)$, \ie\ $\Phi'$, has disappeared from the effective 
metric. A closer look at the apparent gauge field mass term in 
(\ref{Ricci5dim}) shows this to be\ $\int dx^+\del_+(NN')$, which 
vanishes being the integral over a compact direction of a total 
derivative. On the other hand, the scalar kinetic terms do in fact 
contribute a mass term for the gauge field: we have the terms
\be
-{1\over 2} g^{++} (\del_+\Phi)^2 - g^{+t} \del_+\Phi \del_t\Phi\ 
\ra \ -{1\over 2} N^2 (1-A_0^2) (\Phi')^2 + \ldots
\ \ra\ {1\over 2} N^2 (\Phi')^2 A_0^2\ .
\ee
With\ $N^2={1\over\gamma (\Phi')^2}$\ , the mass term becomes\ 
${m_A^2\over 2}={1\over 2\gamma}$, \ie\ $m_A^2=4$ ($AdS_5$) or 
$m_A^2=2$ ($AdS_4$), agreeing with \cite{mcgreevyBalasubLifBH}.

The 5-dim metric is a solution to the Einstein equations with a scalar 
depending only on the $x^+$-direction. Then the $[00]$-component equation 
of motion gives\
${1\over 2} (-6 - {1\over 8} F_{0i}^2) = -4$, which gives\
$(\del_iA_0 + A_0{\del_iK\over K})^2 = {4\over w^2}$. 
admitting the solution\ $K={1\over w^2}\ , A_0=-1$.
These conditions are also satisfied by a massive ($m^2 = 4$) vector 
field with profile $A=A_0e^0=A_0Kdt=-{dt\over w^2}$ in the $z=2, d=2$ 
Lifshitz (bulk) background metric ($Lif_{z=2}^{d=2}$):
\be\label{lifz=2}
ds^2 = -{dt^2\over w^4} + {{dx^i}^2\over w^2} + {dw^2\over w^2}\ .
\ee
As mentioned earlier, the fluxes that source $Lif_{z=2}^{d=2}$ are 
classically equivalent to a massive vector field with profile 
$A=-{dt\over w^2}$\ . As a further check, assuming the scalar is a 
function of $x^+$ alone, the scalar equation of motion simplifies to\
$\del_+ (N^2 (1-A_0^2) {K\over w^3} \Phi') = 0$,\ verifying again 
the above solution. Note that time reversal invariance is broken in 
these solutions, by the gauge field in the lower dimensional system, 
and by the metric in the higher dimensional one.

What we have demonstrated here is that the on-shell Lifshitz spacetime 
with massive gauge field source is a solution to a 5-dim effective 
action corresponding to Einstein gravity with a massive gauge field 
and two scalars, one the remnant of the 10-dim dilaton and the other 
the Kaluza-Klein scalar corresponding to the radius of the compact 
dimension. The on-shell solution relates the two scalars and further 
fixes the gauge field mass in terms of the two scalars.

It is perhaps surprising that the naive dimensional reduction
(\ref{metlifsh}) involves\ $\Phi'\sim {1\over N(x^+)}$\ which however
disappears in the metric (\ref{met4Lif}) implied by (\ref{Ricci5dim}):
we do not have an intuitive way to understand this. The nontrivial
dependence on the $x^+$-dimension might appear to complicate a
Wilsonian separation-of-scales argument making it harder to justify
why it is consistent for modes other than the ones here to be
trivial\footnote {We thank J. McGreevy and S. Trivedi for emphasizing
  this.}. For instance, one could imagine turning on a lower
dimensional vector potential $A_idx^i$: this would arise from a
Kaluza-Klein gauge field 1-form\
$A=A_0e^0+A_ie^i={A_0\over N}{\bar e}^0+A_i{\bar e}^i$, with 
corresponding field strength\ $dA={1\over 2} F_{\mu\nu}e^\mu\wedge e^\nu$.
We do not have any conclusive result here for a consistent dimensional 
reduction: for instance, the 5-dim Ricci scalar has extraneous factors 
of $N(x^+)$ appearing in the analogous calculation, making it harder 
to interpret the lower dimensional system.
However it is tempting to believe that some generalization of our 
``minimal'' Kaluza-Klein ansatz (containing only $A_0$) will address 
these concerns and possibly also pave the way for more general 
Lifshitz spacetimes\footnote{It appears difficult however to find 
more general solutions in the higher dimensional $AdS_5$-deformed 
system within these ansatze or minor generalizations: in particular, 
attempts, in the cosmological context (S. Das, KN, S. Trivedi, 
unpublished), to find solutions with radial dependence for the 
dilaton (and metric) were not conclusive.}.

The calculation for the $AdS_4$-deformed solution is similar, resulting 
in a $2+1$-dimensional bulk $z=2$ Lifshitz theory. In sec. 6, we will 
find an alternative approach to uplift the $Lif_{z=2}^{d=2}$ background 
to 11-D supergravity.

\subsection{Scalar probes and Lifshitz geometry}

We would like to see how a bulk supergravity scalar sees the Lifshitz 
geometry at long wavelengths.\\
Consider the scalar action\ 
$S={1\over G_5} \int d^5x\ \sqrt{-g}\ g^{\mu\nu}\del_\mu\varphi\del_\nu\varphi$:\
on restricting to modes with no $x^+$-dependence (\ie\ 
$\del_+\varphi=0$), this gives 
\bea
S &=& {1\over G_5} 
\int {d^4x dx^+\over w^5} \left[ -{w^4 (\Phi')^2\over 4} (\del_-\varphi)^2 
+ w^2 (\del_i\varphi)^2 + w^2 (\del_w\varphi)^2 \right]\nonumber\\
&& =\int {d^4x \over w^5} \left[ -w^4 \left({\int dx^+ (\Phi')^2\over 4}\right) 
(\del_-\varphi)^2 + w^2 L (\del_i\varphi)^2 + w^2 L (\del_w\varphi)^2 \right]
\nonumber\\
&& = {1\over G_4} \int {d^4x\over w^5} \left[ -w^4 (\del_-'\varphi)^2 
+ w^2 (\del_i\varphi)^2 + w^2 (\del_w\varphi)^2 \right]\ ,
\eea
where $L$ is the size of the compact $x^+$-dimension, and $G_4={G_5\over L}$ 
is the 4-dim Newton constant arising from dimensional reduction.

Thus we see that after the rescaling\ 
$x^-\ra x^{-'}={L\over \int dx^+ (\Phi')^2} x^-$,\ the scalar action at 
wavelengths long compared to the compactification size becomes that in 
the 4-dim $z=2$ Lifshitz background (\ref{lifz=2}).

A priori, this looks slightly different from a direct dimensional reduction 
of the equation of motion of the scalar, where it would seem that $\Phi'$ 
remains. The calculation here suggests that the Lifshitz geometry arises
on scales large compared with the typical scale of variation (\ie\ the 
compactification size), in other words effectively setting 
$\Phi'\sim const$.

\section{The dual field theory}

The field theory dual to the $AdS_5$ backgrounds is the $d=4$ \Nf\ super 
Yang-Mills theory with an appropriate lightlike deformation: taking 
the scalar to be the dilaton, the identification is essentially that 
given in \cite{dmnt1,dmnt2}, \ie\ the \Nf\ SYM theory with the gauge
coupling deformed to vary along the $x^+$-direction as\ 
$g_{YM}^2(x^+)=e^{\Phi(x^+)}$. Note that in the PBH coordinates 
(\ref{metPBH}), (\ref{metPBH2}), the boundary metric\ 
$ds_4^2=\lim_{r\ra 0} r^2 ds_5^2$\ 
on which the gauge theory lives is manifestly flat space. The lightlike 
deformation means that no nonzero contraction exists involving the 
metric and coupling alone, since only $\del_+\Phi$ is nonvanishing 
with $g^{++}=0$: thus various physical observables (in particular the 
trace anomaly) are unaffected by this deformation.\\
In the conformal coordinates (\ref{metconf5}), the base space on which
the gauge theory lives is conformal to flat space with metric\
$\tilde{g}_{\mu\nu}=e^{f(x^+)}\eta_{\mu\nu}$. Various arguments were
given in \cite{dmnt2} discussing the role of the lightlike conformal
factor in the gauge theory. The lightlike nature implies that various
physical observables are in fact unaffected by the conformal factor
since no nonzero contraction exists. However an important role played
by the conformal factor is in providing dressing factors for operators
and their correlators: specifically, conformally dressed operators in 
the conformally flat background behave like undressed operators in
flat space, as we will discuss below in the context of the holographic
2-point function. The gauge coupling is again subject to the lightlike
deformation alone as\ $g_{YM}^2(x^+)=e^{\Phi(x^+)}$.

In lightcone gauge $A_-=0$ (compatible with Lorentz gauge 
$\del_\mu A^\mu=0$), the gauge kinetic terms reduce to those for the 
transverse modes $A_i$, the field $A_+$ being nondynamical: this is 
essentially similar to multiple copies of a massless scalar. Retaining 
modes of the form\ $A_i\equiv e^{ik_+x^+} A_i(x^-,x^i)$, with momentum 
$k_+$ along the $x^+$-direction, and approximating the coupling by its 
mean value say $g_{YM}^{(0)}$, this gives
\be
\int d^3x dx^+ {1\over g_{YM}^2(x^+)}\ 
[-2\del_+A_i \del_-A_i + (\del_j A_i)^2]\ \ra\ 
\int d^3x\ {L\over (g_{YM}^{(0)})^2}\ [-iA_i\del_tA_i + {1\over k_+} 
(\del_j A_i)^2]\ ,
\ee
identifying $x^-\equiv t$, absorbing a $k_+$ into the definition of $A_i$, 
with $L$ being the size of the compact $x^+$-direction. This heuristic 
argument shows the $z=2$ Lifshitz scaling symmetry in the kinetic terms. 
In a sense, this is not surprising, since the $z=2$ Lifshitz symmetry 
can be obtained by breaking Galilean (Schrodinger) symmetries: in the 
present case, the coupling varying along the compact $x^+$-direction 
breaks the $x^+$-shift symmetry. However, the field theory is really 
an interacting strongly coupled field theory with Lifshitz symmetries 
dual to the weakly coupled bulk Lifshitz geometry.\\
After the dimensional reduction along $x^+$, the theory becomes an 
interacting strongly coupled 3-dim gauge theory. The 3-dimensional 
gauge coupling is now naively\ 
${1\over g_{3}^2}=\int dx^+ {1\over g_{YM}^2(x^+)} \sim\ 
{L\over (g_{YM}^{(0)})^2}$ ,\ approximating the 4-dim coupling by its 
mean value. Then the theory is effectively 3-dimensional on length 
scales large compared with the compact direction. 

In a sense, this sort of a DLCQ of \Nf\ SYM with varying coupling is
perhaps better defined than ordinary DLCQ. One would imagine the
coupling variation causes the lightlike circle to ``puff up'',
somewhat akin to momentum along the circle, so that the usual issue 
of strongly coupled zero modes stemming from DLCQ is perhaps less 
problematic here.
This is of course not a rigorous treatment of the dimensional
reduction of the \Nf\ SYM theory, dual to \eg\ the discussion of that
of the bulk metric (\ref{metDimRed}). It would be interesting to
understand this better.

Similarly we expect that the field theory dual to the $AdS_4$
backgrounds is a lightlike deformation, dimensionally reduced, of the 
Chern-Simons theories on M2-branes at supersymmetric singularities 
\cite{abjm,klebanov,terashima,jafferis,imamura,uedayama,lee,ms1,ms2,kleb2,
hanany}, that have been found to be dual to $AdS_4\times X^7$ 
backgrounds, with $X^7$ an appropriate Sasaki-Einstein 7-manifold. 
This is thus a $1+1$-dim field theory. It would also be interesting 
to explore this further.

\subsection{The holographic 2-point function}

The holographic 2-point function of operators ${\cal O}$ dual to 
massive bulk scalars $\varphi$ in this deformed \Nf\ SYM-Lifshitz theory 
can be obtained by the usual rules of AdS/CFT. Doing this calculation
directly in the PBH coordinates (\ref{metPBH}), (\ref{metPBH2}) is 
interesting. However an exact calculation is hindered by the fact that 
the wave equation for a massive scalar does not lend itself to separation 
of variables and solving for the exact mode functions appears difficult:
possible mode functions\ $\varphi(x)=e^{ik_-x^-+ik_ix^i} e^{g(x^+)} \zeta(r)$ 
reduce the wave equation to
\be
-2ik_-g' + {r^3\over\zeta(r)} \del_r \left({1\over r^3} \del_r\zeta(r)
\right) - k_i^2 - {m^2\over r^2} + \gamma r^2 (\Phi')^2 k_-^2 = 0\ ,
\ee
the $r^2 (\Phi')^2$ term being problematic. However, let us consider 
this equation near the boundary $r\ra 0$, where this term is small 
and the metric asymptotes to the $AdS_5$ metric in lightcone 
coordinates. Then one finds the mode functions\ 
$e^{ik_-x^-+ik_ix^i} e^{i(k_i^2-\omega^2) x^+/2k_-} (\omega r)^2 K_\nu(\omega r)$: 
not surprisingly, these are in fact the $AdS_5$ mode functions in 
lightcone coordinates. As we will see below, these also arise in the 
calculation in conformal coordinates (setting $e^f=1$).
This then gives the $AdS_5$ 2-point function in lightcone coordinates
\ $\langle O(x) O(x')\rangle \sim\ 
\frac{1}{[(\Delta{\vec x})^2]^{\Delta}}\ ,$\ 
with $\Delta=2+\sqrt{4+m^2}$.
Note that the distance element arising from the calculation here 
is the 4-dimensional distance\
$(\Delta {\vec x})^2= -2(\Delta x^+)(\Delta x^-) + \sum_{i=1,2} (\Delta x^i)^2$:
this is the analytic continuation of the Euclidean 4-dim distance 
of the boundary theory in pure $AdS_5$.\
Now in the limit of a compactified $x^+$-dimension, with\ 
$\Delta x^+ \ll \Delta x^-, \Delta x^i$, this distance element reduces 
to\ $(\Delta {\vec x})^2\sim\ \sum_{i=1,2} (\Delta x^i)^2$, so that
\be
\langle O(x) O(x')\rangle\ \sim\ {1\over [\sum_i (\Delta x^i)^2]^{\Delta}}\ .
\ee
For a massless bulk scalar, we have $\Delta=4$, recovering the equal time 
2-point function of the (2+1)-dim Lifshitz theory of \cite{KLM}: it also 
corroborates the expectation \cite{ardfendfradkin} that the equal time 
correlators of this (2+1)-dim Lifshitz theory are identical to those of 
a 2-dim Euclidean conformal field theory.

We will now discuss the holographic 2-point function in conformal 
coordinates (\ref{metconf5})  where the conformal factor $e^f$ appears 
explicitly: this calculation has been done in \cite{dmnt2}, noting the 
fact that the scalar wave equation in the lightlike deformed background 
can be solved exactly in these coordinates. We will not repeat this in 
detail here but will describe some essential points. 
Consider a minimally coupled scalar field of mass $m$ propagating in 
the bulk 5-dim metric in (\ref{metconf5}), with action\ 
$S = -\int d^5x \sqrt{-g}\ (g^{\mu\nu}\del_{\mu}\varphi\del_{\nu}\varphi
+m^2\varphi^2)$, that is dual to an operator $O(x)$ in the 
boundary CFT with scaling dimension $\Delta$. The wave equation 
following from the above action can be solved exactly for basis mode 
functions\ $e^{-f(x^+)/2}\ e^{i(k_i^2x^+-\omega^2\int e^f dx^+)/2k_-} 
e^{ik_-x^-+ik_ix^i} (\omega r)^2 K_\nu(\omega r)$,\ where\
$\nu=\sqrt{4+m^2}$.

The scalar action reduces, using the equation of motion, to a term at 
the (regulated) boundary $r=\epsilon$, given as\
$S=-\int d^4x \sqrt{-g} g^{rr}\ \varphi({\vec x},r)\
\del_r\varphi({\vec x},r) |_{r=\epsilon}$ :\ using the basis modes, this 
can be evaluated in momentum space giving (upto an overall 
$\nu$-dependent constant)
\be\label{simpli}
S= \int d^2k_i dk_- dk_+\  \varphi(k_i,k_-, \omega^2) 
\varphi(-k_i,-k_-, \omega^2)\ \omega^{2\nu}\ ,
\ee
where the integrals  over all four variables, $k_i, i=1,2,\ k_-, k_+$ 
go from $[-\infty,\infty]$, and $\omega^2=-2k_-k_+ + k_i^2$. This can 
be recast in position space as
\be\label{posnspAction}
S =C \int d^4x d^4x'\ e^{3f(x^+)/2} e^{3f({x'}^+)/2}\
\varphi({\vec x}) \varphi({\vec x'})\
\biggl(\frac{\Delta\lambda}{\Delta x^+}\biggr)^{1-\Delta}\ \frac{1}{
[(\Delta{\vec x})^2]^{\Delta}}\ ,
\ee
where $C$ is a constant, $\Delta=2+\nu$, and $\lambda=\int e^{f(x^+)} dx^+$\ 
is the affine parameter along null geodesics stretched solely along 
$x^+$. The 4-dimensional distance element here is\
$(\Delta {\vec x})^2= -2(\Delta x^+)(\Delta x^-) + \sum_{i=1,2} (\Delta x^i)^2$.

The boundary coupling between the (boundary value of the) scalar 
$\varphi$ and the operator $O$ is\ 
$S_{Boundary}=\int d^4x \sqrt{-\tilde{g}}\ O(x) \varphi(x)$ ,
where $\tilde{g}_{\mu\nu} =e^f\eta_{\mu\nu} $ is the boundary metric 
and $\varphi(x)=\epsilon^{-\Delta_-}\varphi(x,\epsilon)$, with 
$\Delta_-=2-\nu$.

By the usual prescriptions of AdS/CFT for calculating boundary 
correlation functions, equating the bulk action with the action of the 
boundary theory up to second order in the source $\varphi(x)$ gives
\bea
\sqrt{-\tilde{g}(x)} \sqrt{-\tilde{g}(x')} \langle O(x) O(x') \rangle
&=& \frac{\delta}{\delta\varphi({\vec x})} 
\frac{\delta}{\delta\varphi({\vec x'})}
\langle e^{\int d^4x\sqrt{-\tilde{g}} O(x)\varphi(x)} \rangle_{CFT}
\nonumber\\
&=& \frac{\delta}{\delta\varphi({\vec x})} 
\frac{\delta}{\delta\varphi({\vec x'})}
e^{-S_{Sugra}[\varphi({\vec x})]}\ .
\eea
From (\ref{posnspAction}), we then get
\be
\langle O(x) O(x') \rangle = C e^{-f(x)/2} e^{-f(x')/2}\ 
\biggl(\frac{\Delta\lambda}{\Delta x^+}\biggr)^{1-\Delta}\ \frac{1}{
[(\Delta{\vec x})^2]^{\Delta}}\ .
\ee
It is important to consider correlators of conformally dressed operators
as emphasised in \cite{dmnt2}. For instance, consider the operator $O(x)$ 
above with conformal dimension $\Delta$ in the SYM theory.
Then a simple point to note is that the short distance limit of the 
correlator above gives\ $\langle O(x) O(x')\rangle \sim\ 
e^{-f(x^+)\Delta} {1\over [(\Delta{\vec x})^2]^{\Delta}}$,\ by approximating\ 
${\Delta\lambda\over\Delta x^+}\sim {d\lambda\over dx^+}=e^f$.
Thus it is clear that the conformally dressed operators 
$e^{f(x^+)\Delta/2} O(x)$ have essentially a flat space 2-point function\
$\langle e^{f(x^+) \Delta/2} O(x)  e^{f(x^+) \Delta/2} O(x')\rangle \sim 
{1\over [(\Delta{\vec x})^2]^{\Delta}}$ . In other words, the conformally 
dressed operators in the conformally flat background behave like 
undressed operators in the flat space background. More generally, the 
2-point function for dressed operators at arbitrary points $x, x'$, is
\be
\langle e^{\frac{f(x)\Delta}{2}} O(x) e^{\frac{f(x')\Delta}{2}} O(x')\rangle
= C e^{\frac{f(x)(\Delta-1)}{2}} e^{\frac{f(x')(\Delta -1)}{2}}
\biggl(\frac{\Delta\lambda}{\Delta x^+}\biggr)^{1-\Delta}\ 
\frac{1}{[(\Delta{\vec x})^2]^{\Delta}}\ .
\ee
In the compactified limit, we have $\Delta x^+ \ll \Delta x^-, \Delta x^i$.
It is then consistent to approximate\
${\Delta\lambda\over\Delta x^+}\sim {d\lambda\over dx^+}=e^f$.
Furthermore, it is consistent to approximate $e^{f(x^+)}\sim 1$, 
essentially smearing the $x^+$ dependence relative to the uncompactified 
dimensions. 
This then simplifies the 2-point function for these operators which 
becomes 
\be
\langle e^{\frac{f(x)\Delta}{2}} O(x) e^{\frac{f(x')\Delta}{2}} O(x')\rangle
\ \sim\ \langle O(x) O(x')\rangle\ \sim\ 
\frac{1}{[(\Delta{\vec x})^2]^{\Delta}} \ 
\sim_{\Delta x^+ \ll \Delta x^-, \Delta x^i}\ 
\ {1\over [\sum_i (\Delta x^i)^2]^{\Delta}}\ .
\ee

It is worth noting that the boundary hypersurfaces are different in 
the conformal and PBH coordinates: in the compactified system, they do 
not matter, \eg\ in the 2-point function. Effectively we have smeared 
the conformal factor $e^f\ra 1$. This does not mean that the metrics 
can also be similarly reduced by simply setting $e^f\ra 1$: the radial 
coordinates mix $x^+$-dependence.

\section{$AdS$ time dependent solutions}

With time $t$-dependence rather than lightlike dependence, one has slightly 
more restricted solutions but still a fairly large family (\cite{dmnt1} 
already mentions the AdS Kasner solutions and more can be found in 
\cite{adnt,adnnt}). For instance, the AdS Kasner solutions
\be
ds^2 = {1\over r^2} \left[dr^2 - dt^2 + \sum_i t^{2p_i} (dx^i)^2 
\right] + d\Omega^2\ , \qquad e^\Phi=|t|^\al\ ,
\ee
are nontrivial solutions with the Kasner exponents satisfying 
\be\label{Kasexp}
\sum_ip_i = 1\ ,\qquad \sum_ip_i^2 = 1 - {1\over 2} \al^2\ .
\ee
In this case, the index $i$ ranges over $1,2,3$ and $1,2$ respectively 
for the $AdS_5$ and $AdS_4$ cases, and $d\Omega^2\equiv d\Omega_5^2$ 
or $d\Omega^2\equiv d\Omega_7^2$ respectively. The subfamily with 
$\al=0$, \ie\ trivial dilaton, is nontrivial for the $AdS_5$ case, 
as can be shown by a reparametrization (see \cite{Landau} for a
lucid discussion of these and other anisotropic cosmologies: these
have been discussed more recently in detail in \cite{adnnt}). A
nontrivial dilaton has important consequences: \eg\ it allows the
existence of a symmetric Kasner solution (all $p_i$ equal,
$p_i={1\over 3}$), which is disallowed if $\al=0$, as can be seen from
(\ref{Kasexp}).

These Kasner solutions are seen to admit the following anisotropic 
Lifshitz-like scaling symmetries\footnote{Note that Kasner-like solutions 
with radial $r$-dependence rather than $t$-dependence also exist,
\bea
ds^2 = {1\over r^2} [dr^2 - r^{2p_0} dt^2 + \sum_i r^{2p_i} (dx^i)^2 ] 
+ d\Omega_5^2\ , \qquad e^\Phi=r^\al\ ,\nonumber
\eea
with the conditions\ $p_0+\sum_ip_i=0 ,\ p_0^2+\sum_ip_i^2={\al^2\over 2}$, 
following from the Einstein equations. However, these require a nontrivial 
scalar profile along the radial direction: $\al=0$ forces $p_0,p_i=0$. 
Similar solutions, but without the $AdS$ embedding, have been noted 
in \cite{marikataylor}. After this paper appeared, we were informed of 
\cite{nakayama1}, which notes anisotropic Lifshitz scalings of asymmetric 
Kasner solutions: see also \cite{nakayama2} which studies time-dependent 
deformations of Schrodinger spacetimes.}
\be\label{KasLifsh}
t\ra \lambda t\ ,\qquad r\ra \lambda r\ ,\qquad x^i\ra \lambda^{1-p_i} x^i\ ,
\ee
where $p_i$ are the Kasner exponents above. Although $t,r$, have the 
same scaling, $t,x^i$, have distinct anistropic Lifshitz scaling as one 
would like for the boundary time, space coordinates. This scaling also 
implies a corresponding linear shift of $\log\lambda$ for the dilaton 
$\Phi$ from the scaling $e^\Phi\ra \lambda^\al e^\Phi$. The $AdS_5$-Kasner 
system, as mentioned above, admits nontrivial solutions even with a 
trivial dilaton $\al=0$: these non-dilatonic $AdS_5$-Kasner solutions 
admit true Lifshitz scaling symmetries. This however requires one of 
the exponents $p_i$ to be negative\footnote{\label{Fpi<0}Using 
(\ref{Kasexp}), the exponents $p_i$ can be parametrized as\ 
$p_1=x ,\ p_{2,3}={1\over 2} (1-x\pm\sqrt{1-\al^2+2x-3x^2})$ .\ 
Positivity of the radical requires\ ${1-\sqrt{4-3\al^2}\over 3}\leq x
\leq {1+\sqrt{4-3\al^2}\over 3}$, which for $\al=0$ means\ 
$-{1\over 3}\leq x\leq 1$.\ For $x>0$, we can see from this 
parametrization that $p_1,p_2>0,p_3<0$, while $x<0$ means $p_1<0$.}.\\
Note also that the self-dual 5-form also respects these symmetries. For 
instance, a potentially problematic term\ $*_{10} \omega_5\sim\ 
\sqrt{-g^{(5)}} {dt\wedge dx_1\wedge dx_1\wedge dx_1\wedge dr}$ (where 
$\omega_5$ is the 5-form on $S^5$), is in fact not problematic: 
the scaling of\ $\sqrt{-g^{(5)}}={t^{\sum_ip_i}\over r^5}$\ precisely 
cancels the scaling of the remaining terms.

Besides these $AdS$ Kasner solutions, there are also $AdS_5$-FRW 
solutions \cite{adnt}, one of them with a bounded dilaton. And in fact, 
there is a larger family of scalar 
AdS-BKL cosmological solutions \cite{adnnt} involving $AdS$ embeddings 
of BKL cosmologies \cite{Landau,BKL,BK,misnerBKL}, (see also \cite{Damour}) 
where the spatial metric is one of the homogenous spaces in the Bianchi 
classification (this is discussed at length in \cite{adnnt}, which we 
refer to for details):
\be
ds^2 = {1\over r^2} \left[dr^2 - dt^2 + 
\eta_{ab}(t) (e^a_\alpha dx^\alpha) (e^b_\beta dx^\beta) \right]\ ,
\quad e^\Phi=e^{\Phi(t)}\ ,
\ee
with $e^a_\alpha dx^\alpha$ being a pair of 1-forms defining symmetry 
directions. A spatially homogenous scalar means the spatial $R^a_{(a)}$ 
vanish, and\ $R^0{_0}={1\over 2} (\del_0\Phi)^2$. An interesting 
system here is the $AdS_5$ Bianchi-IX spacetime 
\be
ds^2 = {1\over r^2} \left[dr^2 - dt^2 + 
\eta_i^2(t) e^i_\alpha e^i_\beta dx^\alpha dx^\beta \right] ,
\qquad e^\Phi=|t|^\alpha\ ,
\ee
with three scale factors $\eta_i(t)$. There is an approximate 
Kasner-like solution\ $\eta_i(t)\simeq t^{p_i}$ with\ $\sum_ip_i=1\ ,\ 
\sum_ip_i^2=1-{\alpha^2\over 2}$, if spatial curvatures are ignored.
If all exponents $p_i>0$, the cosmology is ``stable'', in the sense 
that this Kasner regime evolves directly towards the cosmological 
singularity: this is possible only if the dilaton is nontrivial, \ie\ 
$\al\neq 0$ (see Footnote~\ref{Fpi<0}). If initially some $p_i<0$, 
then it turns out that spatial curvatures force a BKL 
bounce\footnote{Let $p_-$ denote a negative Kasner exponent and 
$p_+>0$ being either of the other two positive exponents. Then these 
bounces can be expressed as the iterative map\ 
$p_i^{(n+1)} = {-p_-^{(n)}\over 1+2p_-^{(n)}}\ , \quad p_j^{(n+1)} =
{p_+^{(n)}+2p_-^{(n)}\over 1+2p_-^{(n)}}\ ,$ with\ $\al_{(n+1)} =
{\al_n\over 1+2p_-^{(n)}} ,$\  
for the bounce from the $(n)$-th to the $(n+1)$-th Kasner regime
with exponents $p_i,p_j$. Since $p_-^{(n)}<0$, we have 
$\al_{(n+1)}>\al_{(n)}$, \ie\ the dilaton exponent increases, except 
for the non-dilatonic case $\al=0$.} from one Kasner regime with 
exponents $\{p_i^{(n)}\}$ to a new distinct one $\{p_i^{(n+1)}\}$. 
This oscillatory process continues indefinitely if the dilaton is
trivial. A nontrivial dilaton turns out to drive attractor-like
behaviour, since the dilaton exponent $\al$ increases with each
bounce. The oscillations cease when the system reaches the attractor
basin comprising generic Kasner-like solutions with all $p_i>0$ (see
\cite{adnnt} for details).

Such BKL systems exhibit anisotropic Lifshitz scaling (\ref{KasLifsh}) 
but only approximately since the BKL-Kasner solutions are themselves
only approximate. Incorporating spatial curvatures then means that 
the Lifshitz scaling exponents change as the system bounces from one 
Kasner regime to another.

The gauge theory duals in this case are conjectured \cite{adnnt} to be
the \Nf\ SYM theory living on a time-dependent (and spatially curved)
base space ${\tilde g}_{\mu\nu}$, and with a time-dependent gauge
coupling $g_{YM}^2=e^\Phi$ (in the dilatonic cases). These are highly
non-equilibrium systems with external driving forces (the curved base
spacetime): the rate at which energy is being pumped into the system
is divergent and thus thermalization does not happen, as discussed in
\cite{adnnt}. The time-dependence of the background metric ${\tilde
  g}_{\mu\nu}$ on which the gauge theory lives imparts the
BKL-bouncing behaviour to the gauge theory as well, which is then
forced to bounce from one Lifshitz-regime to another, as time
evolves. If the dilaton (\ie\ the gauge coupling) is constant, then as
we have mentioned, the bounces continue indefinitely towards the
spacelike singularity at $t=0$. From the dual point of view, the
system remains in an approximate Lifshitz regime for some duration,
then is dynamically forced (by the background metric) to bounce to
another, and so on. The bounces themselves are chaotic in the sense
that small perturbations to the initial Kasner-Lifshitz regime give
rise to drastically different subsequent regimes as the bounces occur.
It would be interesting to ask if there are analogous phenomena known
in condensed matter systems, involving smooth transitions between 
regimes of distinct Lifshitz scaling.

It is important to note that these solutions are somewhat different
qualitatively from the null ones described earlier. Most notably, 
they are time-dependent and contain a bulk cosmological 
singularity\footnote{Furthermore there is also a singularity in 
the deep interior ($r\ra\infty$) where the invariant 
$R_{ABCD}R^{ABCD}$ diverges (as does the Ricci scalar): the precise 
gauge theory significance of this is unclear, although one might 
imagine it signals some infrared instability in the gauge theory.} 
at $t=0$. In addition, the bulk in general does not asymptote to 
$AdS_5\times S^5$ at early times\ (the hyperbolic $AdS_5$-FRW with a 
bounded dilaton does though), leaving open the question of initial 
conditions that naturally evolve to the above cosmologies. From the 
boundary point of view, this would mean that the initial state in 
the gauge theory is typically not the vacuum, but some possibly 
non-canonical state. We expect our discussions above pertaining to
\eg\ the BKL bounces will apply once the system lands up in such an
initial state.  Finally it was argued in \cite{adnnt} that the gauge
theory duals in the dilatonic case exhibit a singular response to
these time-dependent deformations of the gauge coupling (in particular
in the symmetric Kasner case, using a PBH transformation to a flat
boundary metric). The theory is likely nonsingular if the coupling 
does not strictly vanish however.

In $AdS_4$, we expect analogs of such cosmological solutions but likely 
with some notable differences. For instance the $AdS_4$ Kasner solutions 
with a trivial scalar ($\al=0$) can be seen to be Milne parametrizations 
of flat space: there are only two exponents $p$ and $1-p$, giving\ 
$p^2+(1-p)^2=1$, \ie\ $p=0,1$. With a nontrivial dilaton, there are 
of course nontrivial cosmological solutions. As another example, the 
$AdS_4$ Bianchi-IX spacetime has a different symmetry algebra, the 
spatial slice spanned by the $e^i$ being only 2-dimensional. The 
system of two Kasner exponents and the scalar one in $AdS_4$ is 
perhaps similar to the non-dilatonic $AdS_5$ Kasner solutions with 
three exponents and the corresponding $AdS_4$ BKL system is perhaps 
oscillatory rather than attractor-like. It would be interesting to 
explore these further.

\section{Further Lifshitz-like solutions in 11-dim supergravity}

Here we consider new solutions in 5-dim gravity with negative
cosmological constant coupled to a massless complex scalar, which are
similar to the null solutions discussed earlier. The $2+1$-dim
Lifshitz spacetimes $Lif_{z=2}^{d=2}$ arise by dimensional reduction
of these 5-dim solutions along one direction. These 5-dim solutions
can be embedded in 11-dim supergravity.

First, we will study a solution in 5-dim with Lifshitz symmetries where 
the shift along $x^+$ is not broken by the metric, but only by a complex
scalar field. The metric and the profile for the complex scalar
field are:
\be
ds^2 = R^2{-2 dx^+ dx^- + d\vec{x}^2 + dw^2 \over w^2} + 
R^2\left({dx^+}\right)^2, 
\quad \varphi(x^+) = \sqrt{2\over \ell^2} {e^{ i\ell x^+}\over R}\ .
\label{5Dsol}
\ee
Here, we have taken the periodicity of $x^+$ to be $2\pi$. The 
normalization of the complex scalar field determines $g_{++}$ and $\ell$ 
is an integer. The background in (\ref{5Dsol}) is an extremum of the 
following action
\be 
S_5 = \kappa_5^2\int d^5 x \sqrt{g_5} \left(R_5 - 2\Lambda - \del_\mu 
\bar{\varphi} \del^\mu {\varphi}\right) \label{5Daction} 
\ee
where $\Lambda = -6/R^2$ and $\bar{\varphi}$ denotes complex conjugate of 
$\varphi$. Note that the 
onshell value of $ \del_\mu \bar{\varphi}\del^\mu {\varphi}$ is zero. This 
fact will be useful in finding an uplift of this solution to 11-D 
supergravity. It is not hard to dimensionally reduce along the 
$x^+$-direction now, as the metric is independent of $x^+$. We will 
use the following ansatz for the line element and the complex scalar 
to perform the KK reduction along $x^+$:
\be
ds^2 = G_{\mu \nu}dx^{\mu} dx^{\nu} + R^2(dx^+ + A)^2, \quad 
\varphi(x^+) = \sqrt{2\over \ell^2} {e^{ i\ell x^+}\over R}\ .
\ee
This generalizes the metric in (\ref{5Dsol}).
The reduced action can be written as
\be 
S_4 = \kappa_4^2\int d^4 x \sqrt{G_4} \left( R_4 - 2 \Lambda -{1\over 4} dA^2 
+ {m^2 \over 2} A^2 \right) 
\ee
where $\Lambda = -6/R^2$ and $m^2 = 4/R^2$. Note that this action is the action 
obtained by dualizing 
the fluxes  in \cite{KLM} as mentioned earlier. Further, the equations of 
motion of the 5-dim action in vielbein indices can be written 
as\footnote{The line element in terms of vielbeins can be written as 
$ds^2 =  \eta_{ab} e^a e^b + e^5 e^5$, where $e^a =  e^a_\mu dx^\mu$, 
$e^5 = dx^+ + A_{\mu} dx^{\mu} = dx^+ + e^5_{\mu} dx^{\mu}$.  Further, 
$dx^\mu = E^{\mu}_a e^a $ and $dx^+ = e^5 - A_a e^a=E^+_{a} e^{a} + 
E^+_{5} e^{5} $. Note that $\del_5{\varphi} = \left(\del_+  + E^{\mu}_5 
\del_\mu \right) \varphi $ and $\del_a{\varphi} = \left(E^+_a\del_+  + 
E^{\mu}_a \del_\mu \right) \varphi$. }
$$ R_{ab} - {1\over 2} R \eta_{ab} - \Lambda\eta_{ab} =  \left(\del_a {\varphi} 
\del_b \bar{\varphi} + h.c \right) - {1\over 2} \eta_{ab} \left(\del_c 
{\varphi} \del^c \bar {\varphi} + \del_5 {\varphi} \del^5 \bar {\varphi} 
\right) $$
$$ \Rightarrow R_{ab} = \left(F_{a c} F^{c}_{b} - {1\over 4} F^2 \eta_{ab} 
+ m^2 A_a A_b\right) - 2{\Lambda} 
\eta_{ab} $$
$$ R_{a 5} = \left( \del_a \varphi \del_5 \bar{\varphi} 
\del_a \varphi \del_5 \bar{\varphi}\right)  \Rightarrow \nabla_a F^{ab} 
= m^2 A^b   $$
$$ R_{55} -{1\over 2} R\eta_{55} - \Lambda\eta_{55} = 2 \del_5 {\varphi} 
\del_5 \bar{\varphi} - {1\over 2} \left(\del_c {\varphi} \del^c \bar 
{\varphi} + \del_5 {\varphi} \del^5 \bar {\varphi} \right) \ \ 
\Rightarrow\  \left( -{1\over 4} F^2 + {m^2 \over 2} A^2 \right) = - \Lambda .$$
Further, the scalar equation of motion is satisfied if and only if
$A^2 = 1$. This condition also guarantees 
$\del_{\mu}\varphi \del^{\mu}\bar{\varphi}=0$. Note that the 5-dim equations 
of motion are satisfied if the 4-dim equations of motion and constraints 
are satisfied. The
$Lif_{z=2}^{d=2}$ metric and matter content solve the above equations
of motion and also satisfies the constraints. Hence, $Lif_{z=2}^{d=2}$
background can be uplifted to a solution of 11-dim supergravity if the
solution in 5-dim can be lifted to a solution of 11-dim supergravity. 
Note that the following eleven dimensional metric and 4-form flux
\be\label{met11dim}
ds^2_{11}  = g_{AB} dx^A dx^B  + ds^2_{\mathbb{CP}^2} + d\chi_1^2 + 
d\chi_2^2~,$$  $$ G_4 = 2 J\wedge J+ 2J\wedge d\chi_1\wedge d\chi_2  + 
\sqrt{3} d\varphi \wedge J \wedge (d\chi_1 - i d\chi_2) + h.c.
\ee
is a solution of 11-dim supergravity if $g^{(5)}$ and $\varphi$ satisfy 
the 5-dim equations of motion along with the constraint
$ d\varphi \wedge_5 \star_5 d\bar{\varphi} = 0$.\footnote{ Using the 
properties of K\"ahler form and the constraint 
$ d\varphi \wedge_5 \star_5 d\bar{\varphi} = 0$, the equations of 
motion of 11-dim supergravity (in the background (\ref{met11dim}))
can be reduced to
$$ R_{\mu\nu} = -4 g_{\mu\nu} + {1\over 2} \left(\del_{\mu}\varphi 
\del_{\nu}\bar{\varphi} + h.c \right),\quad \mbox{for }\mu,\nu = 0,1,2,3 $$
$$ R_{ij} = 6 g_{ij} ,\quad \mbox{for } i,j~ \mbox{in}~ \mathbb{CP}^{2}  $$
All other components of the Ricci tensor vanish. Note that the $i,j$ 
components of Einstein's equations are trivially satisfied. Further, 
the Bianchi identity for the 4-form flux is also trivially satisfied. 
The flux equation is satisfied if $d\star_5 d\varphi = 0$. Hence, the 
conditions for (\ref{met11dim}) to be a solution of 11-dim supergravity 
are the same as the conditions for extremizing the action in 
(\ref{5Daction}).}
Here $\chi_{1,2}$ are coordinates in $S^1\times S^1$ and $J$ is the 
K\"ahler form on $\mathbb{CP}^{2}$. This is similar to some constructions 
in \cite{gauntlettMth}. The $\BC\BP^2$ space here can be generalized to 
any Kahler Einstein space.

Note that $g_{++}$ does not vanish anywhere in this bulk solution. At
this point, we are not clear about the interpretation of the dual
field theory. One might guess that the dual field theory lives on $M5$
branes. Perhaps, it is convenient to study the type II theory on $D4$-
or $D3$-branes obtained by dimensional reduction.

\section{Discussion}

We have discussed $z=2$ Lifshitz geometries obtained by dimensional
reduction along a compact direction of certain lightlike deformations
of $AdS\times X$ solutions of 10- or 11-dimensional supergravity. We
have also described some time-dependent (cosmological) solutions, 
with and without a nontrivial scalar (dilaton), and their anisotropic 
Lifshitz scaling.

Our discussion has been largely from the point of view of the bulk
AdS-deformed theories. The duals in many of these cases are
appropriate deformations of the \Nf\ super Yang-Mills theory. In
particular the constructions in this paper can be taken to suggest
precise field theories dual to AdS-Lifshitz spacetimes. In particular,
the dual to the $z=2$ AdS-Lifshitz theory is simply the dimensional
reduction along the $x^+$-direction of the \Nf\ SYM theory with gauge
coupling $g_{YM}^2=e^{\Phi(x^+)}$. Similarly we expect the $1+1$-dim duals 
in the $AdS_4$-deformed compactified cases are appropriate deformations 
of the Chern-Simons theories on M2-branes at supersymmetric 
singularities. It would be interesting to flesh these out further.

As we have discussed towards the end of sec. 3, the null solutions we
have considered are of a particular type. Generalizing these solutions
with more interesting ansatze, one might expect to find bulk spacetime
solutions describing holographic renormalization group flows between
\eg\ $AdS$ or Schr\"odinger and $z=2$ Lifshitz spacetimes. These would 
correspond to higher dimensional analogs of \eg\ similar RG flows 
discussed in \cite{KLM}. It would be interesting to explore this 
further.

A solution that interpolates between the Schr\"odinger and Lifshitz 
background would break translation symmetry along the $x^{+}$-direction 
in the bulk but not asymptotically. As mentioned earlier, breaking of the 
translation symmetry along the $x^{+}$ direction corresponds to breaking 
the particle number symmetry in the Schr\"odinger spacetime. A solution 
that breaks this symmetry only in the bulk (and not asymptotically), 
describes a state that breaks the particle number (and Galilean boost) 
symmetry spontaneously\footnote{KB thanks J. McGreevy and D. Nickel 
for their discussions on this topic and has been largely motivated by 
these discussions.}. In other words such a solution provides a 
holographic description of a superfluid ground state, in the sense 
that a scalar condensate spontaneously breaks a $U(1)$ global symmetry. 
It would be interesting to explore this further.

\vspace{15mm}
\noindent {\small {\bf Acknowledgments:} It is a great pleasure to thank 
A. Adams, S. Das, J. McGreevy, S. Minwalla, M. Rangamani, V. V. Sreedhar 
and S. Trivedi for useful discussions. We also thank J. McGreevy and 
M. Rangamani for detailed comments on a draft. 
KB thanks the Organizers of the K. S. Krishnan Discussion 
Meeting ``Frontiers in Quantum Science 2009'', held at IMSc, Chennai, 
and the hospitality of CMI where this work began. KN thanks 
the Organizers of the K. S. Krishnan Meeting, IMSc, and of the National 
Strings Meeting NSM09, IIT Bombay for hospitality while this work was 
being carried out. The work of KB is supported in part by funds 
provided by the U.S. Department of Energy (D.O.E.) under cooperative 
research agreement DE-FG0205ER41360. The work of KN is partially 
supported by a Ramanujan Fellowship, DST, Govt of India.}

\vspace{5mm}

\appendix

\section{The general setup for $AdS_5$ cosmological solutions}

The solutions described in (\ref{geomscal}) are in fact part of a more 
general family of solutions of Type IIB supergravity or string theory, 
that are deformations of $AdS_5\times X^5$, with $X^5$ being the base of 
a Ricci-flat 5-dim space. This can be seen by noting that a general 
metric of the form
\be
ds^2 = Z^{-1/2}(x) {\tilde g}_{\mu\nu} dx^\mu dx^\nu
+ Z^{1/2}(x) {\tilde g}_{mn} dx^m dx^n\ , 
\ee
is a solution of the equations of motion, as long as $Z(x)$ is a harmonic 
function on the flat, six dimensional tranverse space with coordinates 
$x^m$, $\tilde{g}_{mn}$ is Ricci-flat, depending only on $x^m$, and
$\tilde{g}_{\mu\nu}$ and the scalar $\Phi$ are dependent only on the 
$x^\mu$, satisfying the conditions (\ref{condns}). Taking the near 
horizon decoupling limit gives the solution in (\ref{geomscal}), the 
$d\Omega_5^2$ now being the metric on the base 5-space over which the 
transverse Ricci-flat space is a cone, with\ 
${\tilde g}_{mn}dx^mdx^n=dr^2+r^2d\Omega_5^2$.

To see how this is obtained, note that the 10D IIB supergravity Einstein 
equations are
\bea
R_{MN}=\frac{1}{6} F_{M A_1A_2A_3A_4}F_N{^{A_1A_2A_3A_4}} 
+ \frac{1}{2}\del_M\phi\del_N\phi\ ,
\eea
the $F^2=F_{ABCDE}F^{ABCDE}$ term vanishing because of the self-duality 
of the 5-form $F$.
For the above backgrounds, it is clear that this equation with components 
along the $S^5$ directions is satisfied, since the scalar does not 
depend on the angular coordinates of the $S^5$: these equations are 
essentially the same as those for the $AdS_5\times X^5$ solution. In 
the $\{\mu,r\}$-directions, the Ricci tensor is
\be
R_{\mu\nu} = {\tilde R}_{\mu\nu} - {4\over R^2} g_{\mu\nu}\ , \qquad 
R_{rr} = -{4\over R^2} g_{rr}\ .
\ee
The term $-{4\over R^2} g_{\mu\nu}$ in the first equation, as well as the 
$R_{rr}$-equation, are balanced by the 5-form contribution (which in 
effect provides a negative cosmological constant in 5-dimensions). 
This shows that the extra contribution ${\tilde R}_{\mu\nu}$ must balance 
the scalar kinetic energy for the Einstein equations with 
$\mu,\nu$-components to be satisfied. In effect the Einstein equation 
then becomes\ $R_{MN}=-4g_{MN}+{1\over 2}\del_M\Phi\del_N\Phi$:\ in fact 
it is easy to see that this equation is also valid when the scalar has 
radial $r$-dependence (as discussed below in the context of $AdS_4$ 
solutions). The scalar equation follows since it satisfies the massless 
free-field equation in 10 dimensions (with a trivial $3$-form field 
strength) and is independent of $r$ and the $S^5$ coordinates.\\
We expect similar solutions exist where the scalar is not the dilaton 
but arises from the 5-form flux through the compactification on a 
nontrivial 5-manifold, as in the $AdS_4\times X^7$ case discussed below.

\subsection{$AdS_4$ null and cosmological solutions}

This is a straightforward generalization to $AdS_4$ of the cosmological 
solutions \cite{dmnt1,dmnt2,adnt,adnnt} described above.

We are considering M-theory backgrounds with nontrivial metric and
3-form, that are generalizations of $AdS_4\times X^7$, with $X^7$ being 
the 7-dim base space (possibly Sasaki-Einstein) of some Ricci-flat 
8-dim space (say a CY 4-fold). With no other matter content, such 
backgrounds can be seen to arise by stacking M2-branes at a point on 
a Ricci-flat transverse space (which is a cone over the 7-dim space 
$X^7$) and taking the near horizon scaling limit, giving the 
$AdS_4\times X^7$ background. 
The 11-dim supergravity equation of motion for the metric components are
\be\label{11dsugraEOM}
R_{MN} = {1\over 12} G_{MB_1B_2B_3} G_N^{B_1B_2B_3} 
- {1\over 144} g_{MN} G_{B_1B_2B_3B_4} G^{B_1B_2B_3B_4} \ ,
\ee
Consider now an ansatz for a deformation of $AdS_4\times X^7$ of the form
\be
ds^2 = {1\over r^2} ({\tilde g}_{\mu\nu} dx^\mu dx^\nu + dr^2) 
+ 4 ds_{X^{7}}^2\ , \quad  
G_{4}=6\mbox{vol} (M_{4}) + C d\Phi(x^\mu)\wedge\Omega_{3}\ ,
\ee
with ${\tilde g}_{\mu\nu}$ being functions of $x^\mu$ alone, the scalar 
$\Phi=\Phi(x^\mu,r)$,\ $C$ being a normalization constant, and 
$\Omega_{3}$ is a harmonic 3-form on some Sasaki-Einstein 7-manifold 
$X^{7}$ with a non-trivial third Betti number ($b_{3}$). With a trivial 
scalar $\Phi=const$ and ${\tilde g}_{\mu\nu}=\eta_{\mu\nu}$, this is 
the $AdS_4\times X^7$ solution (see \eg\ \cite{gauntlettMth} for the 
normalization). The condition $d\Omega_3=0$ ensures 
that the Bianchi identity is satisfied by the 4-form flux, while the 
flux equation $d\star G_4+{1\over 2} G_4\wedge G_4=0$ is satisfied 
if $d(\star d\Phi)=0$ and $d\star\Omega_3=0$: these last two 
equations are the scalar equation of motion and the second condition 
for a harmonic form $\Omega_3$. Further, the Einstein equations for the 
internal indices $i,j$, are satisfied if\ $d\Phi\wedge_4\star_4d\Phi=0 
\sim (\del\Phi)^2$\  (which is consistent with the null solutions 
described in the text). For instance, this kills the scalar terms in the 
second term in (\ref{11dsugraEOM}): further terms involving $\Phi$ in
$G_{iB_1B_2B_3} G_j^{B_1B_2B_3}$ again necessarily force one of the $B_i$ 
to be $\mu$, thus involving the contraction $(\del\Phi)^2$ which 
vanishes. A similar thing is true for the equation with 
$\mu,i$-components, resulting in
$$R_{MN} = -3 g_{MN} + {1\over 2} \del_M\Phi\del_N\Phi\ ,\qquad M,N=\mu,r .$$
In particular, note that this equation also holds for the case when 
the scalar $\Phi$ has radial $r$-dependence. The constant $C$ can be 
used to normalize the coefficient of this scalar kinetic term to be 
${1\over 2}$\ . The 4-form flux provides an effective negative 
cosmological constant in 4-dim.
If $\Phi$ does not depend on $r$, the $rr$-component of this equation 
is simply $R_{rr}=-3g_{rr}$, and the other equations with 
$\mu,\nu$-components simplify to
\be
{\tilde R}_{\mu\nu} = {1\over 2}\del_{\mu}\Phi\del_{\nu}\Phi\ ,\qquad
{} {1\over \sqrt{-{\tilde g}}}\ \del_{\mu} (\sqrt{-{\tilde g}}\ 
{\tilde g}^{\mu\nu} \del_{\nu}\Phi) = 0 ,
\ee
the second equation being the scalar equation of motion. In other words,
a solution to the 3-dim Einstein-scalar system is automatically a solution 
to M-theory on $AdS_4$. It appears difficult to interpret the scalar 
$\Phi$ as the M-theory uplift of the IIA dilaton.

To study time-dependent deformations, we take $\Phi$ and 
${\tilde g}_{\mu\nu}$ to depend only on (i) a time-like variable $t$, or 
(ii) a lightlike variable $x^+$.

\end{document}